\documentclass[twocolumn]{aastex631}

\usepackage{graphicx}
\usepackage{bm}
\usepackage{xcolor}

\begin{document}


\title{Neutron Star Crust Can Support A Large Ellipticity}

\author{J.A. Morales}
\email{jormoral@iu.edu}

\author{C.J. Horowitz}
 \email{horowit@indiana.edu}
 
\affiliation{ Center for the Exploration of Energy and Matter and Department of Physics, Indiana University, Bloomington, IN 47405, USA }


\begin{abstract}
Non-axisymmetrical deformations of the crust on rapidly rotating neutron stars are one of the main targets of searches for continuous gravitational waves.  The maximum ellipticity, or fractional difference in moments of inertia, that can be supported by deformations of the crust (known as ``mountains'') provides an important upper limit on the strength of these continuous gravitational wave sources. We use the formalism developed by Gittins and Andersson, along with a deforming force that acts mainly in the transverse direction, to obtain a maximum ellipticity of 7.4$\times$10$^{-6}$. This is larger than the original results that Gittins and Andersson obtained but consistent with earlier calculations by Ushomirsky, Cutler and Bildsten. This suggests that rotating neutron stars could be strong sources of continuous gravitational waves.
\end{abstract}

\keywords{gravitational waves -- stars: neutron}



\section{Introduction}\label{sec:Intro}

The exciting era of gravitational wave (GW) astronomy began in September 2015, with the detection of the merger of two black holes (BHs) \cite{Abbott2016}. This historic detection introduced a new window to explore the universe. Almost two years later, the LIGO and VIRGO collaborations detected GWs from a binary neutron star (NS) inspiral \cite{Abbott2017}. This was a milestone in GW astronomy and physics, as it was the first detection of GWs coming from NSs, which are some of the most intriguing objects to the GW community.  

GWs have been detected from inspiraling and merging NSs \cite{Abbott2017,Abbott2021}. In addition, NSs can emit continuous gravitational radiation either through r-mode oscillations \cite{Andersson1998} or rotations when there are non-axisymmetrical deformations \cite{Ushomirsky2000,Gittins2021}. In this paper, 
we are interested in non-axisymmetrical deformations of the crust of a NS, which are known as \textit{mountains}.

 Many searches for continuous GWs (CGWs) from NS mountains have been performed. Some of the searches have focused on specific pulsars \cite{Abbott2004,Abbott2005,Abbott2007,Abbott2008,Abbott2010,Abadie2011,Abadie2011_2,Aasi2014,Aasi2015,Aasi2015_2,Abbott2017_2,Abbott2017_3,Abbott2017_4,Abbott2017_5,Abbott2018_2,Abbott2019,Abbott2019_2,Abbott2019_3,Abbott2021Moun}, while others have been wide-parameter surveys for unknown sources \cite{Abbott2005_2,Abbott2007_2,Abbott2008_2,Abbott2009,Abadie2012,Aasi2013,Abbott2016_2,Abbott2017_6,Abbott2018_3,Dergachev2020}. One motivation for these searches is that NSs are known to spin rapidly \cite{Hessels2006}. On the other hand, accreting milisecond pulsars are not spinning near their Kepler frequencies. Since milisecond pulsars do not have strong magnetic fields, \cite{Bildsten1998} suggested that the angular momentum gained from accretion is lost via gravitational radiation. Also, molecular dynamics (MD) simulations have suggested that the crust is likely to be very strong \cite{Horowitz2009,Caplan2018,Kozhberov2020}. We expect many future searches for CGWs with present and next generation GW detectors. 
 
 Mechanisms to produce NS mountains may be complex and uncertain. Therefore an important first step is to calculate the maximum mountain that the crust can support. This maximum NS mountain is useful for two reasons. 
 
 First it provides a bound on the possible strength of CGW sources that is similar to limits from energy conservation or spin down. For example, there have been many searches for CGWs from the Crab pulsar \cite{Abbott_2019a,Abbott2022}. These searches are promising because the Crab is young and energetic. Searches have been able to significantly beat the spin down limit so it is now know that less than 0.02 \% of the spin down power of the Crab is going into CGWs.  However, the Crab spins relatively slowly (about 30 Hz).  Therefore, it takes a large deformation of the crust in order to produce detectable CGWs.  Improved searches may soon beat the maximum mountain limit (i.e., set an observational upper limit on the ellipticity that is smaller than the maximum ellipticity the crust can support). When this is done, CGW searches will be directly probing the shape of the crust of the Crab and setting realistic limits on its deformation. 
 
 In addition, calculating the maximum mountain provides insight into mechanisms that may produce large deformations.  For example, the crust is strongest near its base at high densities.  This region gives a large contribution to the maximum mountain and suggests that mechanisms that build large mountains likely involve the high density inner crust.

\cite{Ushomirsky2000} have calculated the maximum mountain in an intuitive formalism.  They assume the crust can be strained near its breaking strain everywhere and write the maximum deformation the crust can support as a simple integral of the crust breaking stress divided by the local gravitational acceleration.  This yields a maximum ellipticitiy (fractional difference in moments of inertia) of 2.8$\times$10$^{-6}$ for a 1.4 M$_\odot$, 10 km radius NS (see Sec. \ref{subsec::IntroUshomirsky} and Table \ref{Tab:forces}).

Recently, \cite{Gittins2021} developed a novel approach to calculate the maximum mountain that a NS crust can sustain \textit{under the influence of a force}. Their three chosen forces suggested that the maximum elastic deformation that a NS crust can support is approximately $10$ times smaller than the estimate of \cite{Ushomirsky2000}. If this is true, it could reduce our chances to observe CGWs from NS mountains with current and next generation GW detectors \cite{Reed2021}. Nonetheless, these three forces were just examples of how to use their approach. As such, these calculations do not rule out that better forces may be found that give larger deformations. 

In general, an applied force is likely to first break the crust near the surface where the density is low and the crust is very weak. If one limits the force to never break the crust anywhere, this early crust breaking can greatly constrain the strength of the applied force and prevent achieving a large ellipticity. In this paper we avoid breaking the weak outer crust by carefully choosing both the radial and angular form of the applied force. This allows us to use stronger forces and achieve much larger ellipticities.  Our maximum ellipticity of 7.4$\times$10$^{-6}$ is somewhat larger, but consistent, with the maximum ellipticity calculated with \cite{Ushomirsky2000} approach.

\cite{Gittins2021} criticise the calculation of \cite{Ushomirsky2000} because it does not explicitly satisfy boundary conditions at the crust-core interface. To satisfy these boundary conditions, \cite{Gittins2021} use an explicit force to deform a star. Nevertheless, it may be possible to satisfy these boundary conditions with relatively small modifications of the \cite{Ushomirsky2000} calculation, as discussed in Sec. \ref{sec::Pert}.  With optimally chosen forces, the maximum ellipticity may be similar in the \cite{Gittins2021} and \cite{Ushomirsky2000} formalisms. 

This paper is organized as follows. In Sec. \ref{sec::Unpert}, we introduce our background NS model. In Sec. \ref{sec::Pert} we discuss the type of elastic deformations that we consider, as well as the approaches used by \cite{Gittins2021} and \cite{Ushomirsky2000} to calculate the maximum NS mountain. In Sec. \ref{sec::Results}, we present our results for the maximum ellipticity. These are discussed in Sec. \ref{sec::discussion}. We conclude in Sec. \ref{sec::Conclusions} that both the \cite{Ushomirsky2000} and \cite{Gittins2021} formalisms are consistent and they predict that the NS crust can support a large deformation.

Throughout the paper, we use latin letters (like $i$) to refer to spatial indices and primes to refer to radial derivatives. In addition, we use $\delta$ to refer to static Eulerian perturbations. 

\section{\label{sec::Unpert}The Unperturbed Star}

We consider a non-rotating NS with spherical symmetry, a mass of $1.4$ $M_{\odot}$ and a radius of $10$ km, whose matter is in the form of a cold perfect fluid, with mass density $\rho$ and barotropic pressure $p$. We select a polytrope
\begin{equation}
    p(\rho) = K \rho^{1+1/n}
    \label{eqn::EOS}
\end{equation}
as our equation of state, with $n=1$, because it describes well the matter inside massive NSs and it is the equation of state that \cite{Gittins2021} used in their analysis. 

We suppose that the outer $\sim$ $1$ $km$ of the fluid background star solidifies and forms the crust. We assume that the shear modulus of the crust is described by
\begin{equation}
    \mu(\rho) = \kappa \rho
    \label{eqn::SM}
\end{equation}
where $\kappa=10^{16}$ $cm^2 s^{-2}$ \cite{Haskell2006}. This form of the shear modulus facilitates a comparison between our results and \cite{Gittins2021}. Furthermore, we assume that the crust material yields (breaks) when the strain $\bar{\sigma}=\bar{\sigma}_{br}=0.1$ \cite{Horowitz2009}. This is the von-Mises stress criterion applied to the crust of NSs. The breaking strain $\bar{\sigma}_{br}=0.1$ was obtained using MD simulations, and is the breaking strain that \cite{Gittins2021} used in their maximum ellipticity calculations. We assume that the core-crust transition density that marks the base of the crust is $\rho_{base}$ = 2$\times$10$^{14}$ g/cm$^3$. This was the density that both \cite{Ushomirsky2000} and \cite{Gittins2021} used for the base of the crust. \cite{Gittins2021} considered a background NS with a thin fluid ocean between the top of the crust and the surface. They assumed that the crust-ocean transition density that marks the top of the crust is $\rho_{top}$ = 10$^{6}$ g/cm$^{3}$. For numerical convenience, we assume that the crust-ocean interface occurs when the density is $\rho_{top}$ = 2.18$\times$10$^{8}$ g/cm$^3$. Calculations with $10^{7}$ g/cm$^3$ $\leq \rho_{top} \leq 10^{11}$ g/cm$^3$ give similar results. We denote the radius of the bottom interface by $R_1$ and the radius of the top interface by $R_2$ through the rest of the paper.

The Newtonian equations that describe the structure of our solidified background star in equilibrium are the following:
\begin{equation}
m^{\prime} = 4 \pi r^2 \rho \label{eqn::SEa},
\end{equation}
\begin{equation}
p^{\prime} = -\rho \Phi^{\prime} \label{eqn::SEb}, 
\end{equation}
\begin{equation}
\Phi^{\prime} = \frac{Gm}{r^2} \label{eqn::SEc}\, . 
\end{equation}
In these equations, $m(r)$ is the enclosed mass and $\Phi(r)$ is the gravitational potential.
Along with the equation of state (\ref{eqn::EOS}) and the shear modulus (\ref{eqn::SM}), (\ref{eqn::SEa}-\ref{eqn::SEc}) are sufficient to obtain the solidified background star quantities that are needed to make the estimates of the maximum ellipticity that a NS crust can support. 

\section{\label{sec::Pert} The Perturbed Star}

When the star deforms and deviates from perfect sphericity, several mass multipole moments are induced. These moments are related to perturbations in density of the form
\begin{equation}
    \delta \rho_{tot}(\vec{r}) = \sum_{l,m} \delta \rho_{lm}(r) Y_{lm}(\theta,\phi)
    \label{eqn::totalchangedensity}
\end{equation}
where $(l,m)$ denotes the mode of the multipole.
Each moment is defined as
\begin{equation}
    Q_{lm} \equiv \int_0^R \delta \rho_{lm}(r) r^{l+2} dr
    \label{eqn::quadmoment}
\end{equation}
where $R$ is the stellar radius. The ($l$,$m$)=($2$,$2$) mode is the dominant contribution to the CGW signal that a NS mountain emits. Therefore, from now on, we suppose that this is the only moment that the perturbations on the star have. Consequently, we set $l=m=2$ and let $\delta \rho(r) \equiv \delta \rho_{22}(r)$. The ellipticity $\epsilon$ is the fractional difference in moments of inertia,
\begin{equation}
    \epsilon = \frac{I_{xx}-I_{yy}}{I_{zz}}=\sqrt{\frac{8 \pi}{15}} \frac{Q_{22}}{I_0}
    \label{eqn::ellipticity}
\end{equation}
where $I_{zz}=I_0=10^{45}$ g cm$^2$ is the moment of inertia with respect to the axis of rotation of the background star. \cite{Gittins2021} chose this value of $I_0$ to facilitate their comparison of results with observational papers, and it is the one we will use for our calculations as well.

Inspired by these definitions, we will denote any $(l,m)=(2,2)$ first-order perturbation $G$ by
\begin{equation}
    G_{tot}(\vec{r}) = G(r) Y_{lm}(\theta,\phi)\, . 
    \label{eqn::Genpert}
\end{equation}

\subsection{\label{sec::FluidPert} Fluid Perturbations}

For a fluid within the star, it is important to consider both the Euler equation
\begin{equation}
    \nabla_i p + \rho \nabla_i \Phi = 0
    \label{eqn::Euler}
\end{equation}
and the Poisson equation
\begin{equation}
    \nabla^2 \Phi = 4 \pi G \rho\, .
    \label{eqn::Poisson}
\end{equation}
Variations on (\ref{eqn::Euler}) give the perturbed Euler equation,
\begin{equation}
   \nabla_i \delta p_{tot} + \delta \rho_{tot} \nabla_i \Phi + \rho \nabla_i \delta \Phi_{tot} = 0\, .
   \label{eqn::pertEuler}
\end{equation}
Now, suppose that an $(l,m)=(2,2)$ force density of the form
\begin{equation}
    f_i = f_r(r) Y_{lm} \nabla_i r + f_{\perp}(r) r \nabla_i Y_{lm}
    \label{eqn::forcepert}
\end{equation}
acts on the fluid as a first-order perturbation. Then, the perturbed Euler equation (\ref{eqn::pertEuler}) becomes
\begin{equation}
    \nabla_i \delta p_{tot} + \delta \rho_{tot} \nabla_i \Phi + \rho \nabla_i \delta \Phi_{tot} - f_i = 0\, .
    \label{eqn::modpertEuler}
\end{equation}
Also, it is important to consider variations to the Poisson equation (\ref{eqn::Poisson}), which give
\begin{equation}
    \nabla^2 \delta \Phi_{tot} = 4 \pi G \delta \rho_{tot}\, .
    \label{eqn::pertPoisson}
\end{equation}

The perturbed Poisson equation (\ref{eqn::pertPoisson}) gives
\begin{equation}
    \delta \Phi^{\prime \prime} + \frac{2}{r} \delta \Phi^{\prime} - \frac{\beta^2}{r^2} \delta \Phi = 4 \pi G \delta \rho
    \label{eqn::pertPoissonrad}
\end{equation}
where $\beta^2 \equiv l(l+1)$. The angular component of (\ref{eqn::modpertEuler}) gives
\begin{equation}
    \delta \rho = - \frac{1}{c_s^2} ( \rho \delta \Phi - f_{\perp} r )\, .
    \label{eqn::pertdenfluid}
\end{equation}

If the fluid region in consideration contains the center of the star, we require that
\begin{equation}
    \delta \Phi(0) = 0
    \label{eqn::deltaPhireg}
\end{equation}
as we want $\delta \Phi$ to be regular at the center. On the other hand, if the fluid region in consideration contains the surface of the star ($r=R$), then
\begin{equation}
    \delta \Phi^{\prime} (R) + \frac{l+1}{R} \delta \Phi(R) = 0\, .
    \label{eqn::deltaPhiR_fluid}
\end{equation}
\subsection{\label{sec::CrustalPert} Crustal Perturbations}

Strain can build up within the solid crust. The addition of strain in the crust leads to the elastic Euler equation 
\begin{equation}
    \nabla_i p + \rho \nabla_i \Phi - \nabla_j t^{ij}= 0
    \label{eqn::elasticEuler}
\end{equation}
where we suppose that the additional term is a first-order perturbation describing the elastic forces. Therefore, by varying the elastic Euler equation (\ref{eqn::elasticEuler}), we obtain
\begin{equation}
    \nabla_i \delta p_{tot} + \delta \rho_{tot} \nabla_i \Phi + \rho \nabla_i \delta \Phi_{tot} -  \nabla^j t_{ij} = 0\, .
    \label{eqn::elasticEulerpert}
\end{equation}
In the presence of the perturbed force density (\ref{eqn::forcepert}), the perturbed elastic Euler equation (\ref{eqn::elasticEulerpert}) becomes
\begin{equation}
    \nabla_i \delta p_{tot} + \delta \rho_{tot} \nabla_i \Phi + \rho \nabla_i \delta \Phi_{tot} -  \nabla^j t_{ij} - f_i = 0\, .
    \label{eqn::modelasticEulerpert}
\end{equation}
In the last equations, $t_{ij}$ is the shear stress tensor, which is given by
\begin{equation}
    t_{ij} = \mu \left( \nabla_i \xi_j + \nabla_j \xi_i - \frac{2}{3} g_{ij} \nabla_k \xi^k \right)\, .
    \label{eqn::shear-stress-tensor}
\end{equation}
$\vec{\xi}$ is the displacement vector that is suitable for polar perturbations \cite{Ushomirsky2000}, and it is given by the expression
\begin{equation}
    \xi^i = \xi_r(r) Y_{lm} \nabla^i r + \frac{r}{\beta} \xi_{\perp}(r) \nabla^i Y_{lm}\, .
    \label{eqn::static-displacement}
\end{equation}
To make the application of the boundary conditions easier, we can identify the traction vector
\begin{equation}
    T^i = (\delta p_{tot} \delta^{ij} - t^{ij}) \nabla_j r\, .
    \label{eqn::traction}
\end{equation}
\cite{Gittins2021} wrote the traction vector in a form that is even more convenient than that given by (\ref{eqn::traction}):
\begin{equation}
     T^i =[\delta p - T_1] Y_{lm} \nabla^i r - r T_2 \nabla^i Y_{lm}\, .
     \label{eqn::traction2}
\end{equation}

The radial and tangential components of the perturbed elastic Euler equation (\ref{eqn::modelasticEulerpert}), the traction vector (\ref{eqn::traction2}), the displacement vector (\ref{eqn::static-displacement}), and the perturbed Poisson equation (\ref{eqn::pertPoisson}) give the following set of coupled ordinary differential equations (ODEs) \cite{Gittins2021}:
\begin{equation} 
    \xi_r^{\prime} = \frac{1}{r} \xi_r - \frac{\beta}{2r} \xi_{\perp} + \frac{3}{4 \mu} T_1,
    \label{eqn::xi_r_prime}
\end{equation}
\begin{equation} 
    \xi_{\perp}^{\prime} = -\frac{\beta}{r} \xi_r + \frac{1}{r} \xi_{\perp} + \frac{\beta}{\mu} T_2,
    \label{eqn::xi_perp_prime}
\end{equation}
\[
    \left(  1 + \frac{3 c_s^2 \rho}{4 \mu} \right) T_1^{\prime} = \rho \delta \Phi^{\prime} - f_r 
\]
\[             
    - \left[  (c_s^2)^{\prime}( 3 \rho + r \rho^{\prime} ) + c_s^2 \left( \frac{3 \beta^2 \rho}{2 r} + \rho^{\prime} - \frac{r \rho^{\prime 2}}{\rho} + r \rho^{\prime \prime} \right) \right] \frac{1}{r} \xi_r \\
\]
\[            
    + \left[  (c_s^2)^{\prime} 3 \rho + c_s^2 \left(  \frac{3 \rho}{r} + \rho^{\prime} \right) \right] \frac{\beta}{2 r} \xi_{\perp} \\
\]
\[            
     - \left[ \frac{3}{r} + (c_s^2)^{\prime} \frac{3 \rho}{4 \mu} + c_s^2 \left( \frac{3 \rho}{r} - \frac{\rho \mu^{\prime}}{\mu} + \rho^{\prime} \right) \frac{3}{4 \mu} \right] T_1 \\
\]
\begin{equation}           
    + \left( 1 + \frac{3 c_s^2 \rho}{2 \mu} \right) \frac{\beta^2}{r} T_2, \\
    \label{eqn::T_1_prime}
\end{equation}
\[ 
    T_2^{\prime} = \frac{\rho}{r} \delta \Phi - f_{\perp}   
    - c_s^2( 3 \rho + r \rho^{\prime} ) \frac{1}{r^2} \xi_r \\
\]    
\begin{equation}       
    + \left[ \frac{3 c_s^2 \rho}{2} + \left( 1 - \frac{2}{\beta^2} \right) \mu  \right] \frac{\beta}{r^2} \xi_{\perp} 
    + \left(  \frac{1}{2} - \frac{3 c_s^2 \rho}{4 \mu} \right) \frac{1}{r} T_1 - \frac{3}{r} T_2, 
    \label{eqn::T_2_prime}
\end{equation}
\begin{equation} 
    \delta \Phi^{\prime \prime} + \frac{2}{r} \delta \Phi^{\prime} - \frac{\beta^2}{r^2} \delta \Phi = 4 \pi G \delta \rho 
    \label{eqn::z_2_prime}
\end{equation}
where 
\begin{equation}
    \delta \rho = - \left( \frac{3 \rho}{r} + \rho^{\prime} \right) \xi_r + \frac{3 \beta \rho}{2 r} \xi_{\perp} - \frac{3 \rho}{4 \mu} T_1\, .
    \label{eqn::pertdencrust}
\end{equation}

The boundary conditions at either the core-crust transition ($r=R_1$) or the crust-ocean transition ($r=R_2$) are the continuity of both the radial and tangential components of the traction (\ref{eqn::traction2}). These give:
\begin{equation}
    T_{2E} = 0,
    \label{eqn::T2_BC}
\end{equation}
\[
    \rho \delta \Phi_F - f_{\perp F} r = \left( 1 + \frac{3 c_s^2 \rho}{4 \mu} \right) T_{1E} 
\]
\begin{equation}
    + c_s^2 \left[  \left( \frac{3 \rho}{r} + \rho^{\prime} \right) \xi_{r E} - \frac{3 \beta \rho}{2 r} \xi_{\perp E} \right]\, . \\
    \label{eqn::T1_BC}
\end{equation}
In (\ref{eqn::T2_BC}) and (\ref{eqn::T1_BC}), the subscripts $F$ and $E$ refer to the fluid region and the elastic region, respectively. On the surface of the star ($r=R$) the boundary condition is the same as (\ref{eqn::deltaPhiR_fluid}). We assume the density is smooth, so we require the perturbed gravitational potential $\delta \Phi$ and and its radial derivative $\delta \Phi^{\prime}$ to be continuous at the solid-liquid interfaces. Furthermore, the radial displacement $\xi_r$ must be continuous at these interfaces. The same is not necessarily true for tangential displacement $\xi_{\perp}$.

\subsection{\label{subsec::IntroGittins}Gittins et al. calculations }

\cite{Gittins2021} considered background stars similar to those described in Sec. \ref{sec::Unpert}. The background star they considered was a completely fluid star. We'll call it the \textit{fluid star}. The second background star they considered consisted of a fluid core between $0$ and $R_1$, a solid crust between $R_1$ and $R_2$, and a fluid ocean between $R_2$ and $R$. We'll call it the \textit{fluid-solid star}. When both of these background star models were perturbed by the force (\ref{eqn::forcepert}), they described the fluid regions by the ODE (\ref{eqn::pertPoisson}) subject to the boundary conditions (\ref{eqn::deltaPhireg}) and (\ref{eqn::deltaPhiR_fluid}). On the other hand, they described the solid regions by the coupled ODEs (\ref{eqn::xi_r_prime}-\ref{eqn::z_2_prime}) subject to the boundary conditions (\ref{eqn::T2_BC}) and (\ref{eqn::T1_BC}).

To get the maximum ellipticity that a NS crust can support, they considered two maximum ellipticies: the maximum ellipticity of a deformed fluid-solid star and the maximum ellipticity of a deformed fluid star. \cite{Gittins2021} first considered a fluid-solid star. Both the fluid regions and the crust were deformed non-spherically. They varied the amplitude of the force perturbations [finding at the same time the corresponding solutions ($\xi_r$, $\xi_{\perp}$, $T_1$, $ T_2$, $\delta \Phi$, $\delta \Phi^{\prime}$)] until the crust broke at a single point, according to the von-Mises stress criterion, which establishes that the crust yields when the strain $\bar{\sigma} \geq \bar{\sigma}_{br}=0.1$ \cite{Gittins2021,Horowitz2009}. For $(l,m)=(2,2)$ perturbations, \cite{Gittins2021} found that the strain at a single point $(r,\theta,\phi)$ is given by
\[
    \bar{\sigma}^2 = \frac{5}{256 \pi} \bigg\{ 6 \text{sin}^2 \theta \bigg[ 3 \text{sin}^2 \theta \text{cos}^2 2 \phi \bigg( \frac{T_1}{\mu} \bigg)^2  \\
\]
\[
+ 4 ( 3 + \text{cos} 2 \theta - 2  \text{sin}^2 \theta \text{cos} 4 \phi ) \bigg( \frac{T_2}{\mu} \bigg)^2 \bigg] \\
\]
\begin{equation}
     + ( 35 + 28 \text{cos} 2 \theta + \text{cos} 4 \theta + 8 \text{sin}^4 \theta \text{cos} 4 \phi) \left( \frac{\xi_{\perp}}{r} \right)^2  \bigg\} 
    \label{eqn::strain_Gittins}
\end{equation}
This procedure gave them the maximum ellipticity $\epsilon_s$. Then, they considered a completely fluid star and deformed it using the same perturbed force density amplitude that broke one point of the crust of the fluid-solid star, found the corresponding solution ($\delta \Phi$, $\delta \Phi^{\prime}$) and calculated its maximum ellipticity $\epsilon_f$. Finally, they took the absolute value of the difference of both maximum ellipticities, $\Delta \epsilon=|\epsilon_s-\epsilon_f|$, which yields the maximum ellipticity that the crust alone can support, according to their formalism.

To scrutinize \cite{Gittins2021} maximum ellipticity calculation method for an explicit force, we used the following two numerical methods:

\vspace{0.3cm}
\noindent
\textit{Method 1}:  If the force (\ref{eqn::forcepert}) could be written as the gradient of a potential $\chi_{tot}$,
\begin{equation}
    f_i = - \rho \nabla_i \chi_{tot}
    \label{eqn::for_den_pot}
\end{equation}
we grouped together $\delta \Phi_{tot}$ and $\chi_{tot}$ into the variable $U_{tot} = \delta \Phi_{tot} + \chi_{tot}$ (as \cite{Gittins2021} did) so that the ODEs were homogeneous. For instance, \cite{Gittins2021} choose $\chi_{tot} = A r^2 Y_{lm}$ (see equation (\ref{eqn::tidal-like_force})). Then, we generated $n$ arbitrary linearly independent solutions with a fourth-order Runge-Kutta algorithm and found a linear combination whose $n$ constant coefficients were fixed by the $n$ boundary conditions, where $n=6$ for the fluid-solid star and $n=2$ for the fluid star. This method is similar to the one that was used in \cite{Kruger2015}.

\vspace{0.3cm}
\noindent
\textit{Method 2}: If the equations couldn't be written in a homogeneous form, we integrated  the equations with a fourth-order Runge-Kutta algorithm that operated within a globally convergent Newton method (GCNM) \cite{NR}. The GCNM ensured that all of the boundary conditions of both the fluid-solid star and the fluid star were satisfied. 

\subsection{\label{subsec::IntroUshomirsky}Ushomirsky et al. calculations 
}

\cite{Ushomirsky2000} performed the first estimate of the maximum ellipticity that a NS crust can support. They used Newtonian gravity and the Cowling approximation, where the perturbed gravitational potential $\delta \Phi = 0$. The fluid parts were in equilibirum and undeformed with respect to the background star. On the other hand, the crust was deformed with respect to the background star, with deformations that could be described by (\ref{eqn::elasticEulerpert}) [with $\delta \Phi=0$]. Since the only part that was deformed was the crust, the crust was the only part of the star that contributed to the quadrupole moment integral (\ref{eqn::quadmoment}). To perform the integral in the crustal region, they assumed that all of the points of the crust were maximally strained, according to the von-Miss stress criterion. Furthermore, they imposed the vanishing of the shear stresses below and above the crust. They found that a NS crust can sustain a maximum quadrupole moment
\begin{equation}
    Q_{22}^{max} \approx \gamma \bar{\sigma}_{br} I
    \label{eqn::Q22maxU1}
\end{equation}
where $\gamma \equiv 6 \left[  \sqrt{\frac{32 \pi}{15}} + \sqrt{\frac{96 \pi}{5}} \right] \approx 62.13$ is a numerical pre-factor, $g(r)=Gm(r)/r^2$ is the local gravitational acceleration, and 
\begin{equation}
    I \equiv \int_{R_1}^{R_2} \frac{\mu r^3}{g} dr\, .
    \label{eqn::I_Ushomirsky}
\end{equation}
Note that this formula for the maximum quadrupole moment only includes shear stresses on the crust. With the equation of state (\ref{eqn::EOS}) and the shear modulus (\ref{eqn::SM}), we found that (\ref{eqn::Q22maxU1}) reduces to
\begin{equation}
    Q_{22}^{max} \approx 2.2 \times 10^{39} \left( \frac{0.1}{\bar{\sigma}_{br}} \right) g \ cm^2\, ,
    \label{eqn::Q22maxU2}
\end{equation}
for a canonical NS with mass of 1.4 M$_{\odot}$ and radius of 10 km. 

\begin{table}[htb]
\caption{\label{Tab:forces} The maximum ellipticities that we calculated using the approaches of \cite{Ushomirsky2000} and \cite{Gittins2021}. For \cite{Gittins2021} approach, we present different force densities $f_i$. $\epsilon^{s}$ represents the maximum ellipticity for a fluid-solid star, and $\epsilon^{f}$ represents the maximum ellipticity of the corresponding fluid star. $|\epsilon^s - \epsilon^f|$ give the maximum ellipticity that the crust can support, accoriding to \cite{Gittins2021} approach. Note that we reproduced results from \cite{Gittins2021} on the first row. The last three rows contain new results from this work. }
\begin{tabular*}{0.47\textwidth}{c c c c c }
    \hline
    \hline
    Approach 
    & Force & $\epsilon^{f}$ & $\epsilon^{s}$ & $|\epsilon^{s}-\epsilon^{f}|$ \\
    \hline
    \hline
    Gittins & (\ref{eqn::tidal-like_force}) & 3.1$\times$10$^{-2}$ & 3.1$\times$10$^{-2}$ & 2.2$\times$10$^{-8}$  \\
    \hline
    Gittins & (\ref{eqn::force_den_2}) & 6.7$\times$10$^{-5}$ & 6.6$\times$10$^{-5}$ & 6.9$\times$10$^{-7}$ \\
    \hline
    Gittins & (\ref{eqn::force_den}) & 4.7$\times$10$^{-4}$ & 4.7$\times$10$^{-4}$ & 7.4$\times$10$^{-6}$ \\
    \hline
    Ushom. & N/A & N/A & 2.8$\times$10$^{-6}$ & 2.8$\times$10$^{-6}$ \\
\end{tabular*}
\end{table}

\section{\label{sec::Results} Results}

\begin{figure}[tb]
\includegraphics[width=0.90\columnwidth]{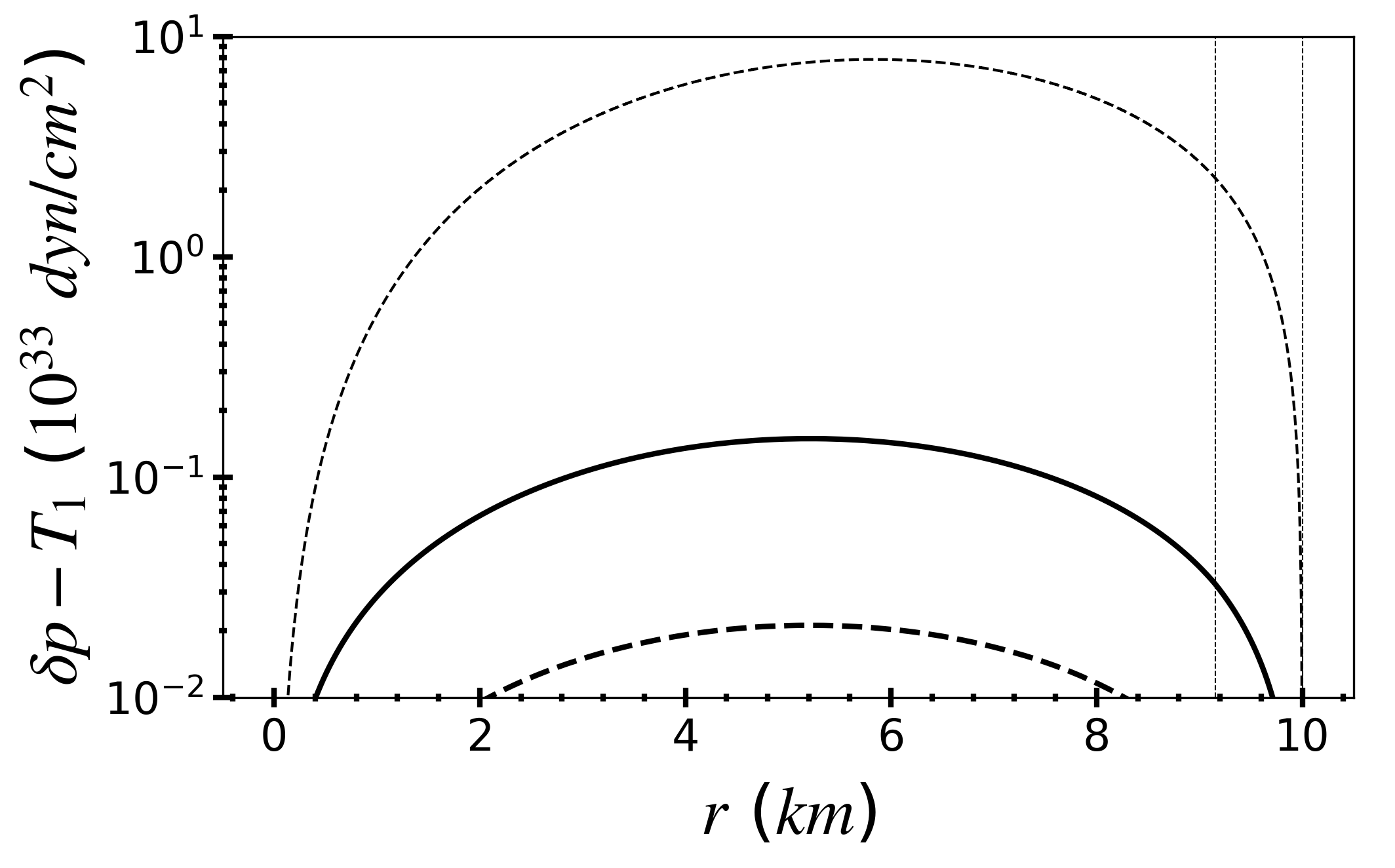}
\caption{Radial component of the perturbed traction as a function of radius for the tidal-like force (\ref{eqn::tidal-like_force}) (thin dashed line), and the forces (\ref{eqn::force_den_2}) (thick dashed line) and (\ref{eqn::force_den}) (solid dashed line).}
\label{fig:dpT1_SM} 
\end{figure}

\begin{figure}[tb]
\includegraphics[width=0.90\columnwidth]{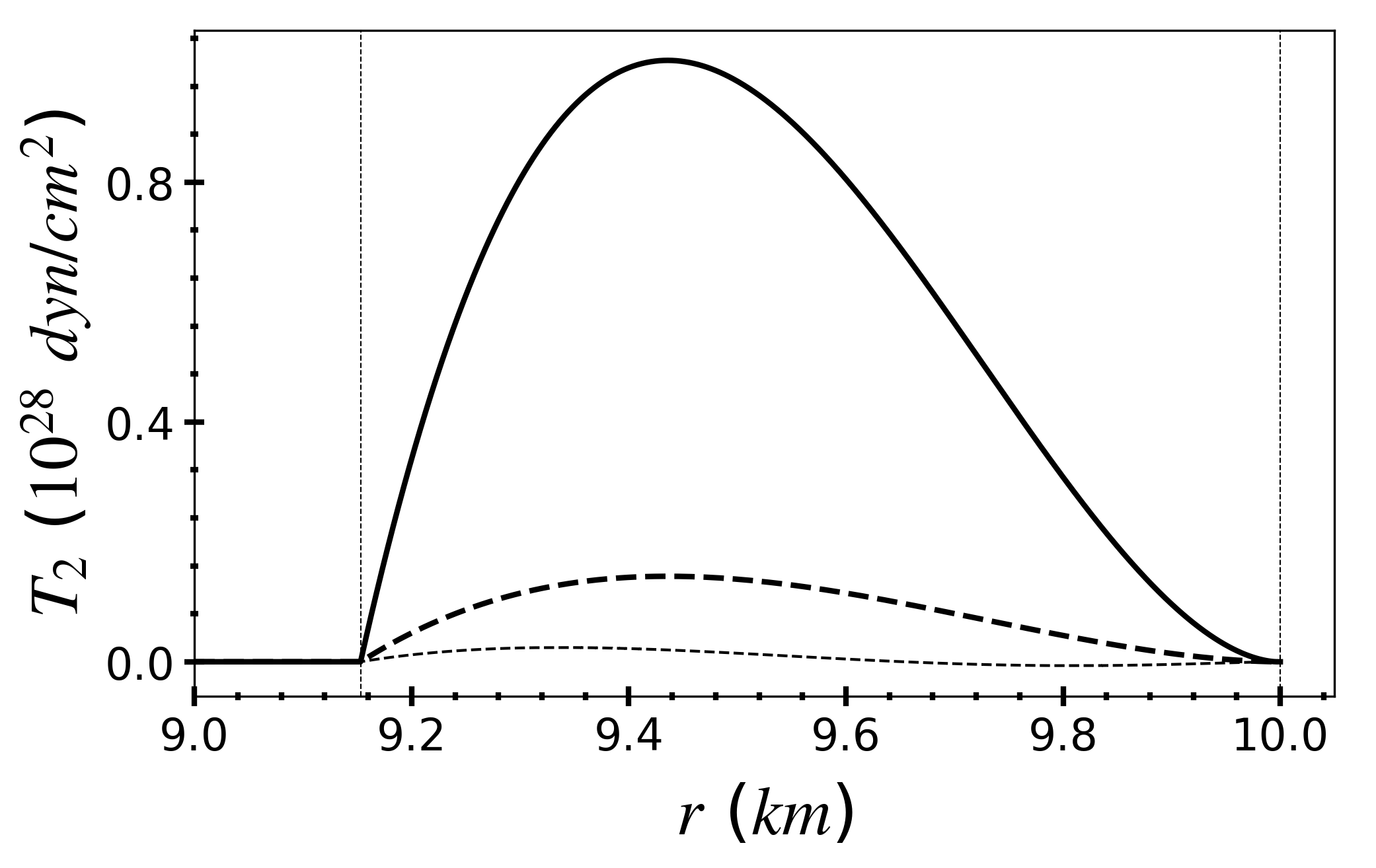}
\caption{Angular component of the perturbed traction as a function of radius for the tidal-like force (\ref{eqn::tidal-like_force}) (thin dashed line), and the forces (\ref{eqn::force_den_2}) (thick dashed line) and (\ref{eqn::force_den}) (solid line).}
\label{fig:T2_SM} 
\end{figure}

\begin{figure}[htb]
\includegraphics[width=0.9\columnwidth]{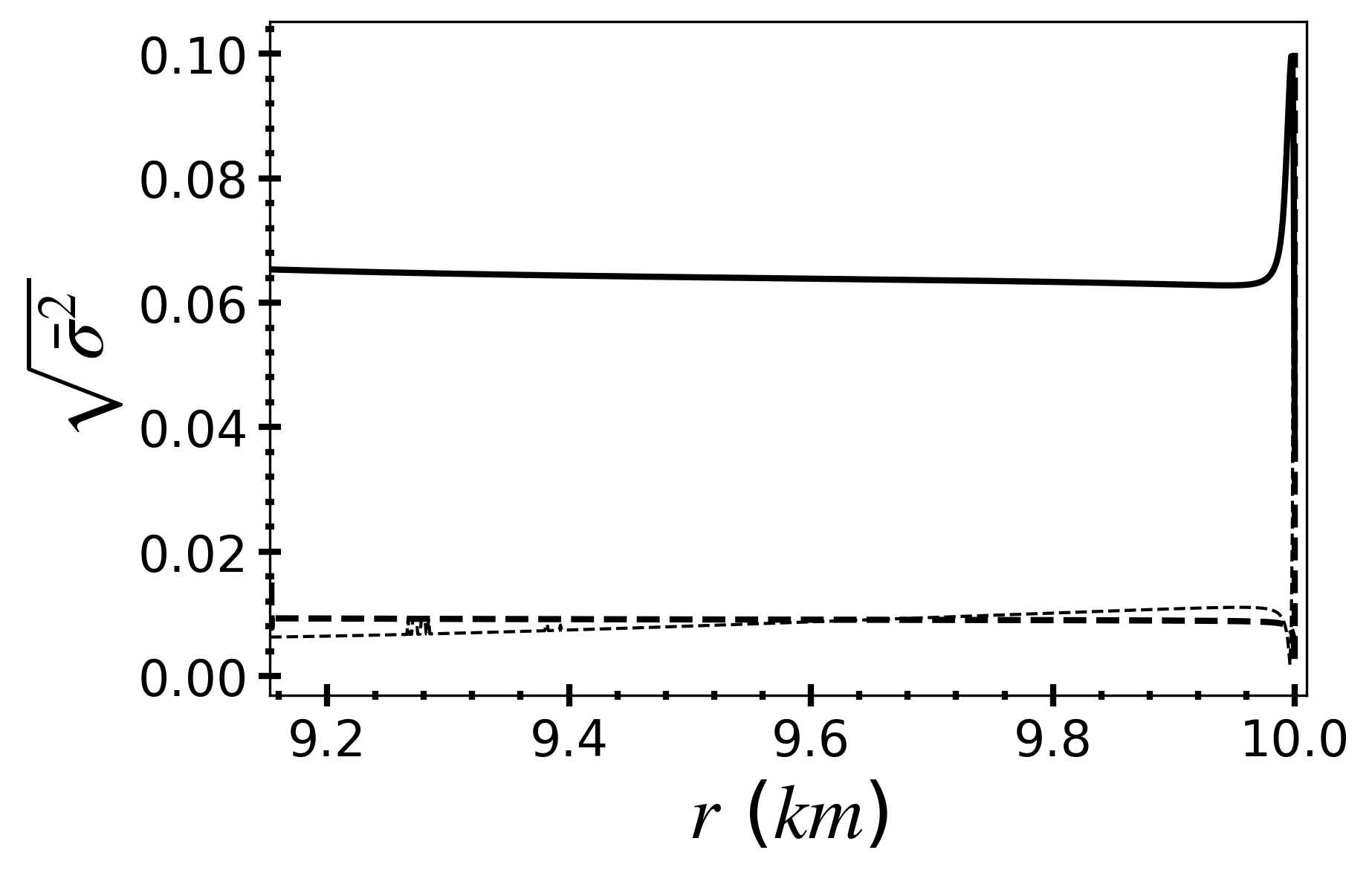}
\caption{Strain as a function of radius for the tidal-like force (\ref{eqn::tidal-like_force}) (thin dashed line), and the forces (\ref{eqn::force_den_2}) (thick dashed line) and (\ref{eqn::force_den}) (solid line).}
\label{fig:strainr_SM} 
\end{figure}

\begin{figure}[htb]
\includegraphics[width=0.9\columnwidth]{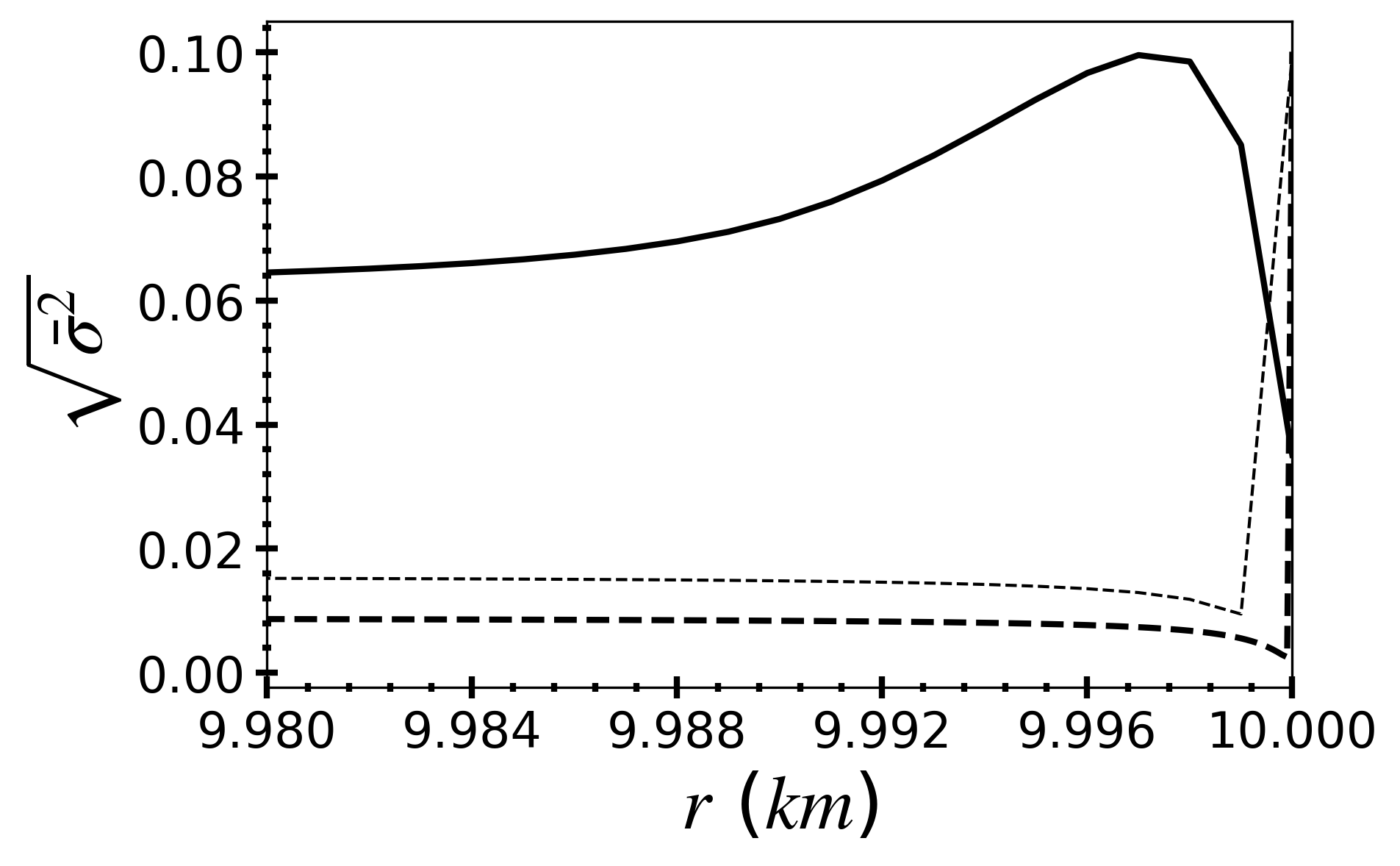}
\caption{Detail of Figure \ref{fig:strainr_SM} for the outer $0.02$ km of the crust.}
\label{fig:strainrshort_SM} 
\end{figure}

In Table \ref{Tab:forces} we collect results for the maximum ellipticity of a NS. We start by reproducing with our code \cite{Gittins2021} results for a tidal-like force 
\begin{equation}
    f_i=-A\rho\nabla_i(r^2Y_{lm})\, .
    \label{eqn::tidal-like_force}
\end{equation}
The radial component of the traction for this force is shown in Figure \ref{fig:dpT1_SM}, while the angular component is shown in Figure \ref{fig:T2_SM}. The amplitude $A$ of the force is adjusted to just break the crust at one point. Unfortunately, this force stresses the very weak outer crust and easily breaks it. This is shown in Figures \ref{fig:strainr_SM} and \ref{fig:strainrshort_SM}. This leads to a strong constraint on $A$ and as a result most of the inner crust is not strained anywhere near the breaking strain. The maximum ellipticity the crust supports in this calculation is as small as $\approx 10^{-8}$ (see Table \ref{Tab:forces}). \cite{Gittins2021} were able to obtain a maximum ellipticity as large as 5.7$\times$10$^{-7}$ with a potential that is a solution to Laplace equation outside of the core. Nevertheless, the corresponding force also stresses and breaks the very weak outer crust. 

\begin{figure}[htb]
\includegraphics[width=0.90\columnwidth]{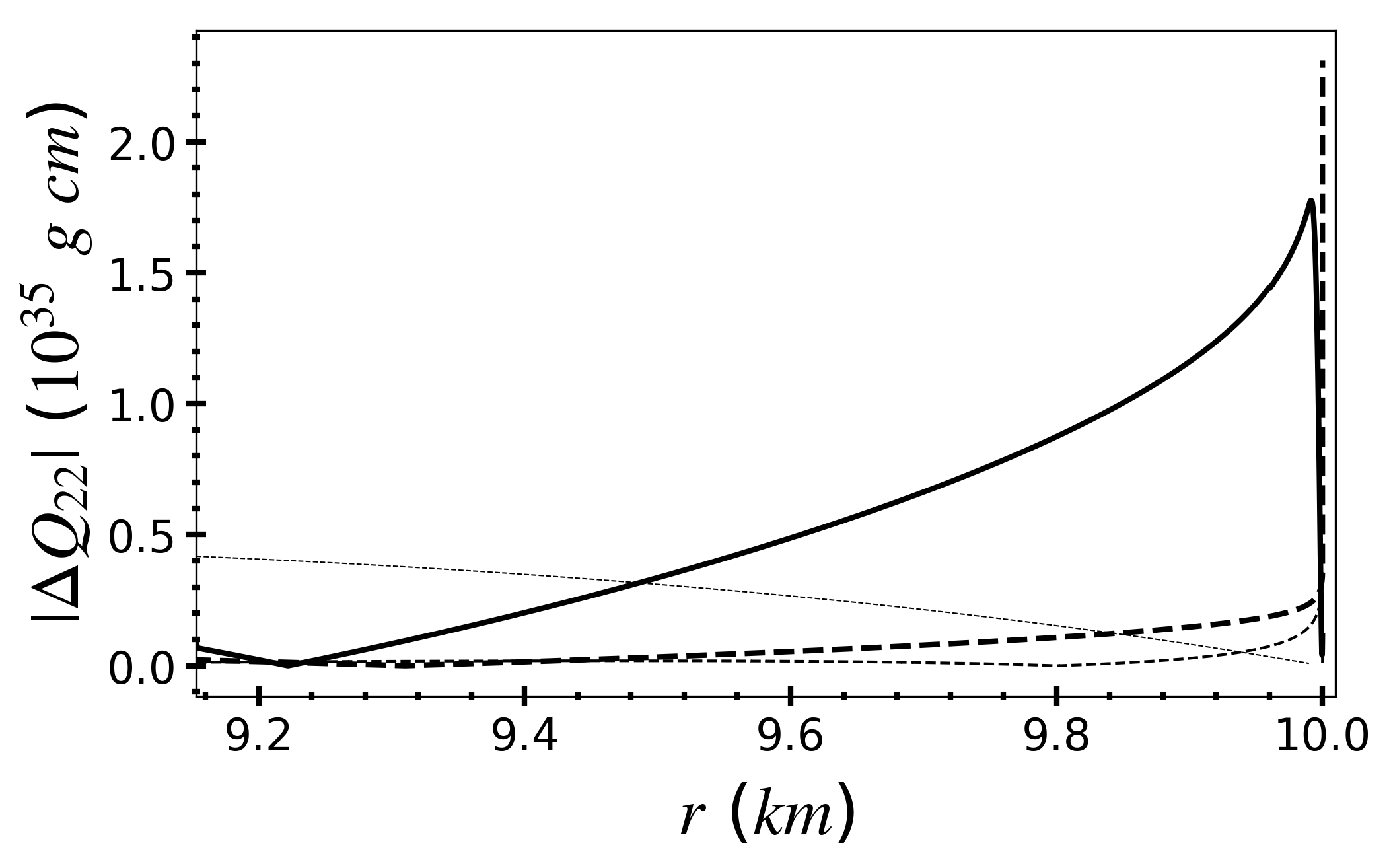}
\caption{Quadrupole moment contribution as a function of radius for the tidal-like force (\ref{eqn::tidal-like_force}) (thin dashed line), the forces (\ref{eqn::force_den_2}) (thick dashed line) and (\ref{eqn::force_den}) (solid line), and \cite{Ushomirsky2000} approach (thin dotted line). The total quadrupole moment is $Q_{22}=\int\Delta Q_{22}(r) dr$.}
\label{fig:dQr_SM} 
\end{figure}

However, one can make better choices for the force $f_i$ that yield much larger ellipticities. We expect $f_i$ that generates a large ellipticitiy satisfies two requirements. First, $f_i$ should strain most the region near the base of the crust because the calculation of \cite{Ushomirsky2000} reveals that the regions near the base of the crust contribute more to the maximum ellipticity when all the crust is maximally strained (see Figures \ref{fig:dQr_SM}). Also, $f_i$ should act primarily in the angular or transverse direction. If a transverse force helps generate a mountain, one can take full advantage of the shear modulus and breaking strain. On the other hand, a radial force will compete with the large incompressibility of the fluid part of the star. As a result elastic forces will only move the mountain a very small radial distance because these forces are quickly counter balanced by a large pressure from a tiny change in the volume of the fluid. 

One might think that force that satisfies these two requirements is one that is proportional to the background density (and, therefore, proportional to the shear modulus within the crust), and acts mainly in the angular direction. Such a force is given by
\begin{equation}
    f_i = B \rho r \nabla_i Y_{lm}\, .
    \label{eqn::force_den_2}
\end{equation}
Nonetheless, this force breaks the crust at the top, leaving most of the crust far from being maximally strained, as shown in Figures \ref{fig:strainr_SM} and \ref{fig:strainrshort_SM}. Therefore, we introduce the force
\begin{equation}
    f_i = B \rho r \nabla_i Y_{lm} + C \Theta (r-r_c) \rho (r-r_c)^2 \nabla_i r Y_{lm}\, .
    \label{eqn::force_den}
\end{equation}
Here $B$ is a constant that is adjusted until the crust breaks at one point, and $r_c=9.99$ km. The constant $C$ is chosen so that we are able to satisfy the additional constraint
\begin{equation}
    T_{1E}(R_2) = 0\, .
    \label{eqn::T1_R2}
\end{equation}
This constraint prevents the crust breaking at its very top and allows a larger $B$ and ellipticity. Note that the extra radial piece in the force (\ref{eqn::force_den}) is turned on smoothly to ensure the density is smooth. 

Figure \ref{fig:dpT1_SM} shows the radial component of the traction for the forces (\ref{eqn::tidal-like_force}), (\ref{eqn::force_den_2}), and (\ref{eqn::force_den}) while Figure \ref{fig:T2_SM} shows the transverse component. As anticipated, the radial traction $\delta p - T_1$ is largest for the tidal-like force (\ref{eqn::tidal-like_force}) (see Figure \ref{fig:dpT1_SM}) and the perpendicular traction $T_2$ is largest for the force (\ref{eqn::force_den}). 
 
The maximum ellipticity for the force (\ref{eqn::force_den}) is given in Table \ref{Tab:forces} as 7.4$\times$10$^{-6}$. This value is even larger than the \cite{Ushomirsky2000} result of 2.8$\times10^{-6}$, but still comparable. Also, note that the maximum ellipticity for the force (\ref{eqn::force_den}) is larger than the maximum ellipticity for force (\ref{eqn::force_den_2}) (refer to Table \ref{Tab:forces}).

\cite{Ushomirsky2000} calculation assumes the Cowling approximation that neglects radial forces. They estimate that relaxing the Cowling approximation results in a maximum ellipticity that is $25-200 \ \%$ bigger than the maximum ellipticity that one can obtain using the Cowling approximation. Our maximum ellipticity for the force (\ref{eqn::force_den}) is $164 \ \%$ larger than what we obtain using \cite{Ushomirsky2000} approach, as can be seen in Table \ref{Tab:forces}. Therefore, our maximum ellipticity of 7.4$\times$10$^{-6}$ is consistent with the estimate of \cite{Ushomirsky2000}. We conclude that the maximum ellipticiity that the NS crust can support is over 10$^{-6}$ in either the \cite{Gittins2021} or \cite{Ushomirsky2000} formalism. A rotating star with a deformation near this value can be a powerful GW source and an attractive target for CGW searches.

\section{\label{sec::discussion} Discussion} 

The keys to obtain a large maximum ellipticity are to first avoid breaking the very weak, low density outer crust and second to strain near its breaking strain the strong, high density inner crust mainly in the angular direction. This allows us to obtain a maximum ellipticity over $10^{-6}$ with either the \cite{Gittins2021} or \cite{Ushomirsky2000} formalism. Although it is true that \cite{Ushomirsky2000} does not explicitly satisfy boundary conditions at the crust-core interface, it is easy to imagine a small modification of their formalism that will. If in a small region near the interface one relaxes the assumption that all of the crust is strained to the breaking strain, one may be able to adjust the strain to satisfy the boundary conditions.

In a second paper, \cite{10.1093/mnras/stab2048} find even smaller maximum ellipticities when they use a shear modulus that is even smaller for the weak outer crust. This result may be misleading. For the limited range of forces that they consider they are not able to strain the inner crust significantly without first breaking the very weak outer crust. With better choices for the form of the force they may be able to produce larger ellipticities. We have not included relativistic effects in this paper. Relativity was found to only modestly reduce the maximum ellipticitiy compared to full nonrelativistic calculations which do not make the Cowling approximation \cite{PhysRevD.88.044004}.

Our large maximum ellipticity provides a useful upper limit to the strength of GWs from rotating NSs. For example the Crab pulsar was formed in a supernova (SN) only one thousand years ago. NS kick velocities indicate that supernovae are energetic and very likely asymmetric at the few percent level. Some of this energy and asymmetry could have been stored in crustal deformations. One mechanism to deform the crust is fall back from the SN. This could have partially melted the newly formed crust on one side and left the NS with a large ellipticity \cite{Janka2022}. Unfortunately, the SN involves complex physics and is difficult to simulate in full detail. Nevertheless, the maximum ellipticity the crust can support still provides a useful upper limit to how large any possible birth deformations can be. Present GW limits for the Crab are not quite at but near this upper limit. As these searches improve and ``beat'' the crust deformation limit their discovery potential may grow and they will directly set realistic limits on the shape of the Crab pulsar.

\section{\label{sec::Conclusions}Conclusions}

The maximum ellipticity that the NS crust can support provides an important limit on the strength of CGWs from rotating NSs. In this work, we have used the \cite{Gittins2021} formalism with an improved deforming force (\ref{eqn::force_den}) to strain the crust mainly in the angular direction in such a way as to produce a maximum ellipticity of $7.4 \times 10^{-6}$. This is larger than the results that \cite{Gittins2021} obtained for three simple forces but consistent with the earlier estimate of \cite{Ushomirsky2000}. This is a promising result for CGW searches because it suggests that rotating NSs could be strong CGW sources.

\begin{acknowledgments}
We would like to thank F. Gittins, N. Andersson and D.I. Jones for their helpful discussions. This work is partially supported by the US Department of Energy, grants DE-FG02-87ER40365 and DE-SC0018083.  
\end{acknowledgments}

\section*{Data Availability}
Additional data for this article will be shared on request to the corresponding author.



\end{document}